\documentclass[aps,preprintnumbers,floats, prd, twocolumn, nofootinbib]{revtex4-1}
\usepackage{graphicx}% Include figure files
\usepackage{dcolumn}% Align table columns on decimal point
\usepackage{bm}% bold math
\usepackage{amsmath}
\usepackage{amsthm}
\usepackage{multirow}
\usepackage{enumerate}
\usepackage{amsfonts}
\usepackage{ifthen}
\usepackage{psfrag}
\usepackage{slashed}
\usepackage{hyperref}
\usepackage{gensymb}
\usepackage[utf8]{inputenc}
\usepackage{color}
\usepackage{subfigure}
\usepackage{ulem}
\usepackage[utf8]{inputenc}

\newcommand{\be}{\begin{equation}}
\newcommand{\ee}{\end{equation}}
\newcommand{\bea}{\begin{eqnarray}}
\newcommand{\eea}{\end{eqnarray}}

\begin{document} 

%\title{Explanation of the XENON1T Excess via Solar Neutrinos in the $U(1)_{B-L}$ Model with a Light Mediator and  Non-Standard Interactions }

\title{The KSVZ Axion and Pseudo-Nambu-Goldstone Boson Models for the XENON1T Excess}

\author{Tianjun Li}
%\email{tli@itp.ac.cn}

\affiliation{ CAS Key Laboratory of Theoretical Physics, Institute of Theoretical Physics, Chinese Academy of Sciences, Beijing 100190, China}
\affiliation{ School of Physical Sciences, University of Chinese Academy of Sciences, Beijing, 100049, China}

\begin{abstract}

The XENON1T excess can be explained by the Axion Like Particle (ALP) dark matter with mass around 2.5~keV. 
However, there are three problems needed to be solved:  suppressing the coupling $g_{a\gamma}$ 
between the ALP and photon, and  generating the proper coupling $g_{ae}$ between the ALP and electron 
as well as the correct ALP mass. We propose three models to solve these problems.
In our models, the $g_{ae}$ couplings are produced by integrating out the vector-like leptons,
and the correct ALP masses arise from high-dimensional operators.
In the KSVZ axion model, the coupling  $g_{a\gamma}$ can be suppressed by choosing
proper sets of vector-like fermions, but we need some fine-tunings to obtain the ALP mass.
Similarly, one can study the DFSZ axion model.
In the  $Z_8$ and $ U(1)_X$ models with approximate Pseudo-Nambu-Goldstone Bosons (PNGBs), 
the coupling  $g_{a\gamma}$  is suppressed due to $SU(3)_C \times U(1)_{\rm EM}$ anomaly free. 
In the  $ Z_8$ model, the PNGB mass can 
be generated naturally at the keV scale via the dimension-8 operator. 
To solve the PNGB quality problem in the $ Z_8$ model,
we embed it into the model with $ U(1)_X$ gauge symmetry.

\end{abstract}

\maketitle

{\bf Introduction.--} Using the low-energy electronic recoil data with an exposure of 0.65 ton-years,
the  XENON Collaboration recently reported the results for new physics search~\cite{Aprile:2020tmw}.
They have oberseved 285 events over an expected background of 232$\pm$15 events,
and found an excess for the electron recoil energies
below 7 keV, rising towards lower energies and prominent between 2 and 3 keV.
Also, they showed that the solar axion and the solar neutrino with magnetic moment can 
provide $3.5\sigma$ and $3.2\sigma$ significance fits to the excess, respectively.
Unfortunately, the correponding parameter spaces are in tension with 
stellar cooling bounds~\cite{Corsico:2014mpa, Giannotti:2017hny, Diaz:2019kim}.
With an unconstrained tritium component in the fitting,
both the solar axion and the solar neutrino magnetic moment
hypotheses no longer have the substantial statistical significance, and their
significance levels are respectively reduced to $2.1\sigma$ and $0.9\sigma$.
This excess has been studied extensively via solar axion, Axion Like Particles (ALPs),
the non-standard neutrino-electron interactions with light mediators, and
dark photon, etc~\cite{Takahashi:2020bpq, Kannike:2020agf, 
Amaral:2020tga, Boehm:2020ltd, Bally:2020yid, AristizabalSierra:2020edu, 
Khan:2020vaf, An:2020bxd, Lindner:2020kko,
DiLuzio:2020jjp, Gao:2020wer, Alonso-Alvarez:2020cdv, Nakayama:2020ikz, Bloch:2020uzh, 
Fornal:2020npv,Chen:2020gcl,Cao:2020bwd,Jho:2020sku, 
Harigaya:2020ckz,Su:2020zny,Lee:2020wmh,Bramante:2020zos,Baryakhtar:2020rwy,
Du:2020ybt,Choi:2020udy,Buch:2020mrg,
Dey:2020sai, Bell:2020bes, Paz:2020pbc, Dent:2020jhf, McKeen:2020vpf, Robinson:2020gfu,
Primulando:2020rdk, DeRocco:2020xdt, An:2020tcg, Ko:2020gdg, DelleRose:2020pbh, 
Chao:2020yro, Bhattacherjee:2020qmv, Gao:2020wfr, Dessert:2020vxy, Alhazmi:2020fju,
Cacciapaglia:2020kbf, Baek:2020owl, Zioutas:2020cul, Sun:2020iim}.

It is well-known that the Peccei-Quinn mechanism~\cite{Peccei:1977hh, Peccei:1977ur} 
provides a natural solution
to the strong CP problem in the Quantum Chromodynamics (QCD), and predicts
a light Pseudo-Nambu-Goldstone Boson (PNGB), dubbed as axion $a$ from  
QCD anomalous $U(1)_{PQ}$ global symmetry breaking. 
The electrwoeak axion~\cite{Peccei:1977hh, Peccei:1977ur, Weinberg:1977ma, Wilczek:1977pj} 
was ruled out by the $K \to  \pi a$ and $J/\Psi \to a \gamma$ experiments. And there
are two viable invisible axion models: 
the Dine-Fischler-Srednicki-Zhitnitsky (DFSZ) model~\cite{Dine:1981rt, Zhitnitsky:1980tq} 
and Kim-Shifman-Vainshtein-Zakharov (KSVZ) model~\cite{Kim:1979if, Shifman:1979if} with
$U(1)_{PQ}$ symmetry breaking scale from about $10^{10}$~GeV to $10^{12}$~GeV.
Interestingly, the ALPs, which are the generalizations of 
axion, may be intrinsic the structure of string theory.
The ALP dark matter can explain the XENON1T excess via the electron 
absorption~\cite{Takahashi:2020bpq, Bloch:2020uzh}, 
and let us study its properties before our model building. 
The Lagrangian between axion and photon/fermions is 
\begin{eqnarray}
\label{eq:Laint1}
\mathcal{L}^{\rm int}_a &\supset& \frac{\alpha_{\rm EM}}{8 \pi} \frac{C_{a\gamma}}{f_a} a F \tilde F
+ C_{af} \frac{\partial_\mu a}{2 f_a} \bar f \gamma^\mu \gamma_5 f 
\, ,
\end{eqnarray}
where $\alpha_{\rm EM}$ is structure constant, $f_a$ is the axion decay constant,
and $C_{a\gamma}$ and $ C_{af}$ are
the couplings. The  above Lagragian can be rewritten as 
\begin{eqnarray}
\mathcal{L}^{\rm int}_a \supset 
 \frac{1}{4} g_{a\gamma} a F \tilde F 
- i g_{af} a \bar f \gamma_5 f ~,~  
\end{eqnarray}
where
\begin{eqnarray}
\label{eq:gagammagaf}
g_{a\gamma} = \frac{\alpha_{\rm EM}}{2 \pi} \frac{C_{a\gamma}}{f_a} \, , \qquad 
g_{af} = C_{af} \frac{m_f}{f_a} 
\, . \nonumber 
\end{eqnarray}
The best fit for the XENON1T excess gives~\cite{Bloch:2020uzh} 
\begin{equation} 
m_a=2.5\text{ keV}~,~~ g_{ae}=2.5\times 10^{-14}~ .~
\label{eq:bestfitALPDM}
\end{equation}
In particular, the cooling constraint $g_{ae} < 2.5 \times 10^{-13}$ 
can be satisfied~\cite{Giannotti:2017hny, Diaz:2019kim}.
The stronger constraint on the decay width for the axion decay into 
diphoton arises from the observation of the cosmic X-ray backgroud (CXB) gives~\cite{Hill:2018trh}  
\begin{eqnarray}
\frac{C_{a\gamma}}{C_{ae}} \lesssim 2.9\times 10^{-3} 
\left(\frac{2.5~{\rm keV}}{m_a}\right)^{3/2}\left(\frac{2.5\times 10^{-14}}{g_{ae}}\right)\,. 
\end{eqnarray}
And then we obtain
\begin{eqnarray}
C_{a\gamma} \lesssim 2.9 \times 10^{-3}  \left(\frac{f_a}{2\times 10^{10}~{\rm GeV}}\right)~.~\,
\label{Cagamma-exp}
\end{eqnarray}
For the QCD axion models, we have
\begin{eqnarray}
C_{a\gamma} &=& \frac{E}{N} -1.92(4).~\,
\label{Cagamma-th}
\end{eqnarray}
where $E$ and $N$ are respectively the electromagnetic and QCD anomaly factors, 
and $1.92(4)$ is generated by the mixing of the axion with
the QCD mesons below the confinement scale.

Next, let us discuss the properites of the ALP dark matter particle, which
can explain the XENON1T excess.
First, we shall show $f_a \simeq 2\times 10^{10}~{\rm GeV}$ later,
and then the traditional QCD axion will have a mass around $2.85\times 10^{-4}$~eV. Thus,
the ALP dark matter particle cannot be the traditional QCD axion.
Second, from Eq.~(\ref{Cagamma-exp}), we obtain $C_{a\gamma} \lesssim 2.9 \times 10^{-3}$.
 In general, there exists about $0.1\%$ fine-tuning for Eq.~(\ref{Cagamma-th}),
and the natural solution to it is 
that both the first term and the second term on the right-handed side vanish:
the first condition implies that we do not have  $[U(1)_{\rm EM}]^2U(1)_{PQ}$
anomaly, while the second condition means no mixing between axion and QCD mesons
and thus we do not have $[SU(3)]^2U(1)_{PQ}$ anomaly.
Therefore,  the ALP dark matter particle, which
can explain the XENON1T excess, might arise from breaking of a $SU(3)_C\times U(1)_{\rm EM}$ 
anomaly free $U(1)_X$ symmetry (or its discrete subgroup) and is a PNGB. 

In short, to explain the XENON1T excess via a PNGB dark matter,
we need to address three problems: how to suppress the coupling $g_{a\gamma}$,
and how to generate the coupling $g_{ae}$ as well as the correct ALP mass.
We shall propose three models to solve these problems: the KSVZ axion model with $U(1)_{PQ}$ symmetry, 
the  model with $Z_8$ discrete symmetry, and the model with  $ U(1)_X$ gauge symmetry.
In our models, assuming that the right-handed electron is charged under $U(1)_{PQ}$, $Z_8$, 
and $U(1)_X$ symmetries, we can produce
the $g_{ae}$ couplings by integrating out the vector-like leptons.
In the KSVZ axion model, the coupling  $g_{a\gamma}$ can be suppressed by choosing
proper sets of vector-like fermions. And with some fine-tuning, we can obtain 
the correct axion mass from high-dimensional operators via quantum gravity effects.
Similarly, one can study the DFSZ model, where the coupling $g_{ae}$ is present and thus 
we do not need to generate it.
In the  $Z_8$ and $ U(1)_X$  models,
 we do not have $SU(3)_C \times U(1)_{\rm EM}$ anomaly, so the coupling  $g_{a\gamma}$  is suppressed.
In the  $SU(3)_C\times SU(2)_L\times U(1)_Y\times Z_8$  model, we obtain the decay constant
 around $2\times 10^{10}$~GeV for the best fit. 
The correct PNGB $a$ mass around 2.5 keV can be generated from dimension-8 operator naturally.
We also show that  $a$ has a lifetime long enough to be a dark matter candidate. Moreover,
the PNGB dark matter density around the observed value can be generated via the misaligment mechanism,
while its thermal density is negligible. Furthermore, to solve the PNGB quality problem via
quantum gravity effects in the $ Z_8$ model, we embed it into 
a $SU(3)_C\times SU(2)_L\times U(1)_{Y}\times U(1)_{X}$ model.
The $U(1)_X$ gauge symmetry is broken down to a $Z_8$ discrete symmetry around the string scale $10^{17}$~GeV,
and then the $Z_8$ model can be realized.

\begin{table}[t]
\begin{tabular}{|c|c|c|c|}
%\hline
% &  $SU(3)_C \times SU(2)_L \times U(1)_Y \times U(1)_{B-L}$& &  $SU(3)_C \times SU(2)_L \times U(1)_Y \times U(1)_{PQ}$\\
\hline
~$XQ_i$~ & ~$(\mathbf{3}, \mathbf{2}, \mathbf{1/6}, \mathbf{1})$~ &
$XQ_i^c$ &  ~$(\mathbf{\overline{3}}, \mathbf{2}, \mathbf{-1/6}, \mathbf{1})$ ~\\
\hline
~$XU_i$~ & ~$(\mathbf{3}, \mathbf{1}, \mathbf{2/3}, \mathbf{1})$~ &
$XU_i^c$ &  ~$(\mathbf{\overline{3}}, \mathbf{1}, \mathbf{-2/3}, \mathbf{1})$ ~\\
\hline
~$XD_i$~ & ~$(\mathbf{3}, \mathbf{1}, \mathbf{-1/3}, \mathbf{1})$~ &
$XD_i^c$ &  ~$(\mathbf{\overline{3}}, \mathbf{1}, \mathbf{1/3}, \mathbf{1})$ ~\\
\hline
$XL_i$ & ~$(\mathbf{1}, \mathbf{2},  \mathbf{-1/2}, \mathbf{1})$~  &
$XL_i^c$ &  $(\mathbf{1}, \mathbf{2},  \mathbf{1/2}, \mathbf{1})$ ~\\
\hline
$XE_i$ & ~$(\mathbf{1}, \mathbf{1},  \mathbf{-1}, \mathbf{1})$~  &
$XE_i^c$ &  $(\mathbf{1}, \mathbf{1},  \mathbf{1}, \mathbf{1})$ ~\\
\hline
~$S$~ &  $(\mathbf{1}, \mathbf{1},  \mathbf{0}, \mathbf{-2})$~
& &   \\
\hline
\end{tabular}
\caption{The particles and their quantum numbers under 
the $SU(3)_C \times SU(2)_L \times U(1)_Y \times U(1)_{PQ}$ gauge and global symmetries. }
\label{Particle-Spectrum-Axion}
\end{table}

{\bf The KSVZ Axion Model.--} First, we construct the KSVZ axion model
which can explain the XENON1T excess. We introduce the vector-like fermions
$(XQ^c_i, ~XQ^c_i)$, $(XU^c_i, ~XU^c_i)$,  $(XD^c_i, ~XD^c_i)$, $(XL^c_i, ~XL_i)$, and $(XE^c_i, ~XE^c_i)$,
as well as a SM singlet axion field $S$.  For simplicity, we assme the vector-like fermions 
have $U(1)_{PQ}$ charge $+1$, while $S$ has $U(1)_{PQ}$ charge $-2$
These particles and their quantum numbers under the
  $SU(3)_C \times SU(2)_L \times U(1)_Y \times U(1)_{PQ}$ gauge and global symmetries 
are summarized in Table~\ref{Particle-Spectrum-Axion}.

The Lagrangian is given by
\begin{eqnarray}
  -{\cal L} &=&  -m_S^2 |S|^2 +\lambda_S |S|^4 
+ \left(y^{XQ}_{ij} S XQ_i XQ_j^c 
\right.  \nonumber\\&& \left.
+ y^{XU}_{ij} S XU_i XU_j^c + 
y^{XD}_{ij} S XD_i XD_j^c \right. \nonumber\\&&\left.
 + y^{XL}_{ij} S XL_i XL_j^c + y^{XU}_{ij} S XE_i XE_j^c
 + {\rm H. C.}\right)~.~\,    
  \label{lag:yukawa}
\end{eqnarray}

To have small $C_{a\gamma}$, 
we need to find the sets of vector-like fermions which gives $E/N$ close to 1.92(4). 
Because the contribution to the electromagnetic anomaly factor from $(XL^c_i, ~XL_i)$
is the same as the $(XE^c_i, ~XE^c_i)$, we do not consider $(XE^c_i, ~XE_i)$ for simplicity.
Of course, any $(XL^c_i, ~XL^c_i)$ can be replaced by a $(XE^c_i, ~XE_i)$ in the following discussions.
For $n$ pairs of $(XQ^c_i, ~XQ^c_i)$, $m$ pairs of $(XU^c_i, ~XU^c_i)$,
$k$ pairs of $(XD^c_i, ~XD^c_i)$, and $l$ pairs of  $(XL^c_i, ~XL^c_i)$,
we obtain the condition  $C_{a\gamma}\simeq 0$
\begin{eqnarray}
\frac{10n+8m+2k+6l}{6n+3m+3k}&\simeq&1.92(4)~.~\,
\end{eqnarray}
It is not difficult to find the approximate solution to the above equation, for example
$\frac{10n+8m+2k+6l}{6n+3m+3k}=2$ for $n=m=0$, $k=6$, and $l=4$.
In addition, assuming that the right-handed electron and muon are charged under $U(1)_{PQ}$ symmetry
and introducing the vector-like fermions $(XL_1,~XL_1^c)$ and $(XL_2,~XL_2^c)$, we can generate 
the coupling $g_{aee}$ as we discuss in the following  $ Z_8$ and $ U(1)_X$ models. 
If the QCD axion only obtains mass via instanton effect, its mass will be too small
since the decay constant is around $10^{10}$~GeV as in the following discussions.
Therefore, the key question is how to generate the correct axion mass around $2.5$~keV.
As we know, the global $U(1)_{PQ}$ symmetry can be broken by the quantum gravity effects.
To be concrete, we consider the following effective operator with dimension $d = 2m + n$
that violates the PQ symmetry by $n$ units~\cite{Irastorza:2018dyq}
\begin{align} 
\label{eq:PQbreakMpl}
 V & \supset 
\frac{ \lambda^m_n  |S|^{2m} \left(e^{-i\delta^m_n} S^n+ e^{i\delta^m_n }{S^\dagger}^n\right)}{{\rm M_{Pl}}^{d-4}}  
\nonumber \\
%& \supset   \frac{ \lambda_n  f^4_a}{2} \! 
% \left(\!\frac{f_a}{\sqrt{2} {\rm M_{Pl}}}\!\right)^{\!d-4}\cos\left(\frac{na}{f_a} - \delta^m_n\right) \nonumber \\
&\approx  m^2_{*}f^2_a\left(\frac{\theta^2}{2}  -\frac{\theta}{n} \tan\delta^m_n\right) \, ,\nonumber
\end{align}
where we have expanded for  $\theta = \frac{a}{f_a}\ll1$ by neglecting an irrelevant constant.
Here, $M_{\rm Pl}$ is the reduced Planck scale, $\lambda^m_n$ is real and $\delta^m_n$ the phase of the coupling,
 $S = \frac{1}{\sqrt{2}} (f_a + s) e^{i a/f_a}$, and 
$m^2_{*} =  \frac{ \lambda^m_n f^2_a}{2}  \left(f_a / (\sqrt{2} {\rm M_{Pl}} )\right)^{d-4}\!\!\cos\delta^m_n $.
In particular, the linear term or tadpole term will shift the QCD vacuum from $\langle \theta \rangle =0$.
Therefore, if we have multiple high-dimensional operators, we can find the fine-tuned solution
where the sum of the linear terms is zero or so small that the solution to the strong CP problem
can be preserved. And the condition is
\begin{eqnarray}
\sum_{m,n} \frac{\tan\delta^m_n}{n} &\simeq& 0~.~\,
\end{eqnarray}
Also, the axion mass is given by
\begin{eqnarray}
m_a&=& 
\sqrt{\sum_{m,n} \frac{ \lambda^m_n f^2_a}{2}  \left(f_a / (\sqrt{2} {\rm M_{Pl}} )\right)^{d-4}\!\!\cos\delta^m_n} ~.~\,
\end{eqnarray}
Therefore, with some fine-tuning, we have shown that the KSVZ axion model can explain the XENON1T excess.
Similarly, one can study the DFSZ model, where the coupling $g_{ae}$ is present, and then 
we do not need to generate it.

{\bf The $Z_8$ Model.--} We shall propose a $SU(3)_C\times SU(2)_L\times U(1)_{Y}\times Z_8$ model
where $Z_8$ is a global discrete symmetry. First, let us explain our convention, which is the same
as the supersymmetric Standard Model (SM). The SM quark doublets, right-handed 
up-type quarks, right-handed down-type quarks, lepton doublets, right-handed charged leptons,
 right-handed neutrinos, and the SM Higgs doublet are denoted as 
$Q_i$, $U_i^c$, $D_i^c$, $L_i$, $E_i^c$, $N_i^c$, and $H$, respectively. 
We shall construct the models where  the masses and mixings for the SM quarks and neutrinos
are generated in a traditional way.
Thus, $Q_i$, $U_i^c$, $D_i^c$, $L_i$, $N_i^c$, and $H$ are not charged under $Z_8$ discrete
 symmetry. Also, we assume that the $Z_8$ quantum numbers for right-handed electron $E_1^c$, muon $E_2^c$, 
and tau $E_3^c$ are $+1$, $-1$, and $0$, respectively. To break the $Z_8$ gauge symmetry 
and have a approximate PNGB, we introduce a SM singlet scalar $S$ with charge $-1$ under $Z_8$. 
Moreover, to generate the electron and mun Yukawa couplings, we introduce two pairs of vector-like fermions 
$(XL_1,~XL_1^c)$ and $(XL_2,~XL_2^c)$. 
These particles and their quantum numbers under the
  $SU(3)_C \times SU(2)_L \times U(1)_Y \times Z_8$ gauge and discrete symmetries are summarized
in Table~\ref{Particle-Spectrum}.

\begin{table}[t]
\begin{tabular}{|c|c|c|c|}
%\hline
% &  $SU(3)_C \times SU(2)_L \times U(1)_Y \times U(1)_{B-L}$& &  $SU(3)_C \times SU(2)_L \times U(1)_Y \times U(1)_{B-L}$\\
\hline
~$Q_i$~ & ~$(\mathbf{3}, \mathbf{2}, \mathbf{1/6}, \mathbf{0})$~ &
$U_i^c$ &  ~$(\mathbf{\overline{3}}, \mathbf{1}, \mathbf{-2/3}, \mathbf{0})$ ~\\
\hline
~$D_i^c$~ & ~$(\mathbf{\overline{3}}, \mathbf{1}, \mathbf{1/3}, \mathbf{0})$ 
&~$L_i$~ & ~$(\mathbf{1}, \mathbf{2},  \mathbf{-1/2}, \mathbf{0})$~ ~\\
\hline
$E_1^c$ &  $(\mathbf{1}, \mathbf{1},  \mathbf{1}, \mathbf{1})$ &
$E_2^c$ &  $(\mathbf{1}, \mathbf{1},  \mathbf{1}, \mathbf{-1})$ ~ ~\\
\hline
$E_3^c$ &  $(\mathbf{1}, \mathbf{1},  \mathbf{1}, \mathbf{0})$ &
~$N_i^c$~ &  $(\mathbf{1}, \mathbf{1},  \mathbf{0}, \mathbf{0})$~ \\
\hline
$XL_1$ & ~$(\mathbf{1}, \mathbf{2},  \mathbf{-1/2}, \mathbf{-1})$~  &
$XL_1^c$ &  $(\mathbf{1}, \mathbf{2},  \mathbf{1/2}, \mathbf{1})$ ~\\
\hline
$XL_2$ & ~$(\mathbf{1}, \mathbf{2},  \mathbf{-1/2}, \mathbf{1})$~  &
$XL_2^c$ &  $(\mathbf{1}, \mathbf{2},  \mathbf{1/2}, \mathbf{-1})$ ~\\
\hline
$H$ & ~$(\mathbf{1}, \mathbf{2},  \mathbf{-1/2}, \mathbf{1})$~  &
~$S$~ &  $(\mathbf{1}, \mathbf{1},  \mathbf{0}, \mathbf{-1})$~   \\
\hline
\end{tabular}
\caption{The particles and their quantum numbers under 
the $SU(3)_C \times SU(2)_L \times U(1)_Y \times Z_8$ gauge and discrete symmetries. }
\label{Particle-Spectrum}
\end{table}

The scalar potential in our model is given by
\begin{eqnarray}
 V &=& -m_S^2 |S|^2 -m_H^2 |H|^2 +\lambda_S |S|^4 
+ \lambda_{SH} |S|^2 |H|^2  \nonumber\\&&
+ \lambda_H |H|^4 + \frac{y}{M^4_{\rm Pl}} |S|^8 
+ \frac{1}{M^4_{\rm Pl}}\left( y' S^8 
+{\rm H.C.}\right) ~.~
\label{scalar-potential}
\end{eqnarray}
 For simplicity, we assume $y > |y'|$ so that the potential
is stabilized. From the the dimension-8 operator 
$y' S^8/M^4_{\rm Pl} $,
we obtain the mass of the PNGB $a$  is 
at the order of $|\langle S \rangle|^6/M_{\rm Pl}^4$.

The Lagrangian for the Yukawa couplings and vector-like fermion masses is
\begin{eqnarray}
  -{\cal L} &=& y_{ij}^U Q_i U_j^c \overline{H} + y_{ij}^D Q_i D_j^c H + y_{i3}^E L_i E_3^c H
  + y_{ij}^{\nu} L_i N_j^c \overline{H}  \nonumber\\&&
+y^{XL}_1 XL_1 E_1^c H + y^{XL}_2 XL_2 E_2^c H + y^S_i S L_i XL_1^c
\nonumber\\&&
+ y^{\overline{S}}_i  \overline{S} L_i XL_2^c + M^N_{ij} N_i^c N_j^c + 
M^{XL}_1 XL_1 XL_1^c 
\nonumber\\&&
+ M^{XL}_2 XL_2 XL_2^c
 + {\rm H. C.}~,~\,    
  \label{lag:yukawa}
\end{eqnarray}
Using $y_{ij}^{\nu} L_i N_j^c \overline{H}$ and $M_{ij}^{N} N_i^c N_j^c$ terms,
we can generate the neutrino masses and mixings via Type I seesaw mechanism.
For simplicity, we choose $y^S_1 \not=0$ and $y^{\overline{S}}_2=0$,
while $y^S_{2} = y^S_{3} =y^{\overline{S}}_1=y^{\overline{S}}_3= 0$.
After integrating out the vector-like particles $(XL_1, XL_1^c)$ and $(XL_2, XL_2^c)$, we obtain
\begin{eqnarray}
-{\cal L} &\supset& -\frac{1}{M^{XL}_1} y_1^{XL} y_{1}^{S} S L_1 E_1^c H 
\nonumber\\&&
-\frac{1}{M^{XL}_2} y_2^{XL} y_{2}^{\overline{S}} \overline{S} L_2 E_2^c H
 + {\rm H. C.}~.~\,
\end{eqnarray}
Thus,  we obtain
\begin{eqnarray}
f_a ~\equiv~ \langle S \rangle &=& \frac{m_e}{g_{ae}} ~=~2\times 10^{10}~{\rm GeV} \times \left(\frac{2.5\times 10^{-14}}{g_{ae}}\right)
~.~\, \nonumber 
\end{eqnarray}
Therefore, for the best fit, we have $f_a =2\times 10^{10}$~GeV. And
then the mass of the PNGB $a$ is around keV scale
from the dimension-8 operator $y' S^8/M^4_{\rm Pl} $ 
in Eq.(\ref{scalar-potential}),
and we can indeed take it as 2.5 keV.    

After integrating out the electron and muon, we obtain
the effective Lagrangian between the PNGB $a$ and photon~\cite{Nakayama:2014cza} 
\begin{eqnarray}
 {\cal L}_{\rm eff} &=& \frac{\alpha_{\rm em} m_a^2}{48 \pi f_a} 
 \left(\frac{1}{m_e^2} - \frac{1}{m_{\mu}^2}\right)
 a F_{\mu \nu} \tilde{F}^{\mu \nu} \, . 
 \label{ALP-photon}
\end{eqnarray}
And then we get
\begin{eqnarray}
C_{a\gamma} &=& \frac{1}{6} \left(\frac{m_a^2}{m_e^2}-\frac{m_a^2}{m_{\mu}^2}\right)
\simeq 4.17\times 10^{-6}~,~\,
\end{eqnarray}
which is much smaller than $2.9\times 10^{-3}$ and is negligible.
The PNGB $a$ can decay into two photons via the above effective interaction,
and the decay rate is~\cite{Nakayama:2014cza} 
\begin{eqnarray}
 && \Gamma_{a \to \gamma \gamma} 
 ~\simeq~ 
 \frac{\alpha_{\rm em}^2 q_e^2}{9216 \pi^3} \frac{m_a^7}{m_e^4 f_a^2} \,
 \nonumber\\
 &&\simeq~ 4.17 \times 10^{-57} \, {\rm GeV}~ 
 \left(\frac{m_a}{2 ~{\rm keV}}\right)^7 
\left(\frac{2\times 10^{10} ~{\rm GeV}}{f_a} \right)^{2} \, . 
 \nonumber
\end{eqnarray}
So the constraint on the flux of the X-ray photons produced by the PNGB decay
can be satisfied~\cite{Takahashi:2020bpq, Caputo:2019djj}. 

In our model, the relativistic PNGBs can be produced from 
the scatterings between electron/muon and the Higgs bosons in the thermal bath. 
The resulting abundance is~\cite{Nakayama:2014cza} 
\begin{eqnarray}
 \Omega_{a}^{\rm (th)}  h^2 \sim && 3.28\times 10^{-4}
 \left( \frac{T_R}{3 \times 10^5 ~{\rm GeV}} \right)
 \left( \frac{m_a}{2.5 ~{\rm keV}} \right) \nonumber\\
 &&
\times \left(\frac{2\times 10^{10} ~{\rm GeV}}{f_a} \right)^{2} ~,~\, 
 \label{Omega_ALP_th}
\end{eqnarray}
where $T_R$ is the reheating temperature. Thus, the thermal relic density of $a$ is
negligible.

The PNGB $a$ can be produced  by the misalignment mechanism as well. 
When the Hubble parameter is smaller than the mass of $a$, 
 it begins to oscillate around its potential minimum. 
The temperature $T_{\rm osc}$ at the onset of the PNGB oscillation  is~\cite{Takahashi:2020bpq, Nakayama:2014cza}  
\begin{eqnarray}
 T_{\rm osc} \sim 1.12\times 10^6 ~{\rm GeV} \left( \frac{m_a}{2.5~{\rm keV}} \right)^{1/2} \, .
\end{eqnarray}
For the temperature higher than $T_{\rm osc}$, the PNGB field $a$ has a field value 
which is not the potential minimum in general. 
We define the initial oscillation amplitude as $a_{\rm Initial} \equiv \theta_{\rm mis} f_a$
with $\theta_{\rm mis}$ the misalignment angle,  and obtain
the oscillation energy of the PNGB $a$~\cite{Takahashi:2020bpq, Nakayama:2014cza}
\begin{eqnarray}
 \Omega_{a}^{\rm (mis)} h^2 \sim 
 && \, 0.1
 \left(\frac{\theta_*}{4}\right)^2
 \left( \frac{f_a}{2\times 10^{10} ~{\rm GeV}} \right)^{2} 
 \nonumber\\
 && \times
 \left\{ 
 \begin{array}{ll}
 \left( \frac{T_R}{10^6 ~{\rm GeV}} \right) \quad &{\rm for}~~ T_R \lesssim T_{\rm osc}
 \vspace{0.1cm}\\
 \left( \frac{m_a}{2.5 ~{\rm keV}} \right)^{1/2} 
 \quad &{\rm for}~~T_R \gtrsim T_{\rm osc}
 \end{array}
 \right. \, .
\end{eqnarray}
Thus, to realize the observed dark matter relic density, we need large 
initial misalignment angle. We consider the reheating temperature
is higher than the oscillation temperature, and the $Z_8$ symmetry breaking
is after inflation. Thus, the decays of the topological defects 
such as cosmic string and domain wall might contribute to the relic 
density of the PNGB $a$ as well.

{\bf The $U(1)_X$ Model.--} In the above model,
 $Z_8$ is a discrete symmetry, and can be broken via the quantum gravity effects.
Thus, the above discussions might not be valid in general if we consider quantum gravity corrections, 
which is called the PNGB quality problem. Because we do not solve the strong CP problem, in principle 
we are fine with quantum gravity corrections since we can fine-tune some parameters in our models.
To solve the PNGB quality problem, we propose the
$SU(3)_C \times SU(2)_L \times U(1)_Y \times U(1)_X$ model where
the $U(1)_X$ gauge symmetry is broken down to the $Z_8$ discrete symmetry around the string scale $10^{17}$~GeV.
In addition to the particles in the $Z_8$ model, we shall introduce two pairs of 
vector-like particles $(XE_1, XE^c_1)$ and  $(XE_2, XE^c_2)$, as well as 
a SM singlet Higgs scalar field $T$ with $U(1)_X$ charge 8.
The particles and their quantum numbers under the
  $SU(3)_C \times SU(2)_L \times U(1)_Y \times U(1)_X$ gauge symmetry are given
in Table~\ref{Particle-Spectrum-UIX}. And one can easily show that
our model is anomaly free.

\begin{table}[t]
\begin{tabular}{|c|c|c|c|}
%\hline
% &  $SU(3)_C \times SU(2)_L \times U(1)_Y \times U(1)_{B-L}$& &  $SU(3)_C \times SU(2)_L \times U(1)_Y \times U(1)_{B-L}$\\
\hline
~$Q_i$~ & ~$(\mathbf{3}, \mathbf{2}, \mathbf{1/6}, \mathbf{0})$~ &
$U_i^c$ &  ~$(\mathbf{\overline{3}}, \mathbf{1}, \mathbf{-2/3}, \mathbf{0})$ ~\\
\hline
~$D_i^c$~ & ~$(\mathbf{\overline{3}}, \mathbf{1}, \mathbf{1/3}, \mathbf{0})$ 
&~$L_i$~ & ~$(\mathbf{1}, \mathbf{2},  \mathbf{-1/2}, \mathbf{0})$~ ~\\
\hline
$E_1^c$ &  $(\mathbf{1}, \mathbf{1},  \mathbf{1}, \mathbf{1})$ &
$E_2^c$ &  $(\mathbf{1}, \mathbf{1},  \mathbf{1}, \mathbf{-1})$ ~ ~\\
\hline
$E_3^c$ &  $(\mathbf{1}, \mathbf{1},  \mathbf{1}, \mathbf{0})$ &
~$N_i^c$~ &  $(\mathbf{1}, \mathbf{1},  \mathbf{0}, \mathbf{0})$~ \\
\hline
$XL_1$ & ~$(\mathbf{1}, \mathbf{2},  \mathbf{-1/2}, \mathbf{-1})$~  &
$XL_1^c$ &  $(\mathbf{1}, \mathbf{2},  \mathbf{1/2}, \mathbf{1})$ ~\\
\hline
$XL_2$ & ~$(\mathbf{1}, \mathbf{2},  \mathbf{-1/2}, \mathbf{1})$~  &
$XL_2^c$ &  $(\mathbf{1}, \mathbf{2},  \mathbf{1/2}, \mathbf{-1})$ ~\\
\hline
$XE_1$ & ~$(\mathbf{1}, \mathbf{1},  \mathbf{-1}, \mathbf{-1})$~  &
$XE_1^c$ &  $(\mathbf{1}, \mathbf{1},  \mathbf{1}, \mathbf{0})$ ~\\
\hline
$XE_2$ & ~$(\mathbf{1}, \mathbf{1},  \mathbf{-1}, \mathbf{1})$~  &
$XE_2^c$ &  $(\mathbf{1}, \mathbf{1},  \mathbf{1}, \mathbf{0})$ ~\\
\hline
$H$ & ~$(\mathbf{1}, \mathbf{2},  \mathbf{-1/2}, \mathbf{1})$~  &
~$S$~ &  $(\mathbf{1}, \mathbf{1},  \mathbf{0}, \mathbf{-1})$~   \\
\hline
$T$ &  $(\mathbf{1}, \mathbf{1},  \mathbf{0}, \mathbf{8})$ & & \\
\hline
\end{tabular}
\caption{The particles and their quantum numbers under 
the $SU(3)_C \times SU(2)_L \times U(1)_Y \times U(1)_X$ gauge symmetry. }
\label{Particle-Spectrum-UIX}
\end{table}

The scalar potential  is given by
\begin{eqnarray}
 V &=& -m_S^2 |S|^2 -m_T^2 |T|^2
-m_H^2 |H|^2+\lambda_S |S|^4 
\nonumber\\&&
+\lambda_T |T|^4 
+\lambda_H |H|^4
+ \lambda_{ST} |S|^2 |T|^2  
\nonumber\\&&
+ \lambda_{SH} |S|^2 |H|^2
+ \lambda_{TH} |T|^2|H|^2 
\nonumber\\&&
+ \frac{y}{M^6_{\rm Pl}} |T^2||S|^8 
+ \frac{1}{M^5_{\rm Pl}}\left( y' TS^8 
+{\rm H.C.}\right) ~.~
\label{scalar-potential-UIX}
\end{eqnarray}
To stabilize the potential after $U(1)_X$ gauge symmetry breaking, 
we require 
\begin{eqnarray}
 \frac{y}{M^6_{\rm Pl}}|\langle T \rangle|^2 ~>~ \frac{1}{M^5_{\rm Pl}}|y'\langle T \rangle|~.~\,
\end{eqnarray}

The Lagrangian for the Yukawa couplings and vector-like fermion masses is
\begin{eqnarray}
  -{\cal L} &=& y_{ij}^U Q_i U_j^c \overline{H} + y_{ij}^D Q_i D_j^c H + y_{i3}^E L_i E_3^c H
  + y_{ij}^{\nu} L_i N_j^c \overline{H}  \nonumber\\&&
+y^{XL}_1 XL_1 E_1^c H + y^{XL}_2 XL_2 E_2^c H 
+ y^{XE}_{ik} L_i XE^c_k H \nonumber\\&&
+ y^S_i S L_i XL_1^c 
+ y^{\overline{S}}_i  \overline{S} L_i XL_2^c 
+ y^{\prime \overline{S}}_k \overline{S} XE_1 XE_k^c 
\nonumber\\&&
+  y^{\prime S}_i S XE_2 XE_k^c 
+ M^N_{ij} N_i^c N_j^c + 
M^{XL}_1 XL_1 XL_1^c 
\nonumber\\&&
+ M^{XL}_2 XL_2 XL_2^c
 + {\rm H. C.}~,~\,    
  \label{lag:yukawa-UIX}
\end{eqnarray}
where $i,~j=1,~2,~3$, and $k=1,~2$.

We assume that $T$ acquires a Vacuum Expectation Value (VEV) around string scale
$10^{17}$ GeV, and then the $U(1)_X$ gauge symmetry
is broken down to a discrete $Z_8$ symmetry. To realize the Lagrangian 
in Eq. (\ref{lag:yukawa}), we require the Yukawa couplings $y^{XE}_{ik}$ to be zero or very small.
This can be done in two ways. First,
we introduce a $Z_2$ symmetry under which $(XE_1, XE^c_1)$ and  $(XE_2, XE^c_2)$ are odd
while all the other particles are even. So, the $y^{XE}_{ik} L_i XE^c_k H$ terms will be forbidden.
Because $(XE_1, XE^c_1)$ and  $(XE_2, XE^c_2)$ cannot decay into the SM particles completely
and they are charged particles, we need to require that the reheating temperature is
smaller than their masses, for example, around $10^{10}$~GeV.
Second, we consider the five-dimensional space-time on $S^1/Z_2$, and assume that
the $SU(3)_C \times SU(2)_L \times U(1)_Y \times U(1)_X$ gauge bosons, $S$, $T$,
$(XE_1, XE^c_1)$ and  $(XE_2, XE^c_2)$ are in
the bulk, while all the rest particles are on the 3-brane at $y=0$. In addition,
we assume that the wave functions for $(XE_1, XE^c_1)$ and  $(XE_2, XE^c_2)$ are highly
suppressed on the 3-brane at $y=0$, and then 
the Yukawa couplings $y^{XE}_{ik}$ will be very small.
The rest discussions are similar to the above Section, so we shall not repeat it here.
In short, we can sovle the PNGB quality problem in 
the $SU(3)_C \times SU(2)_L \times U(1)_Y \times U(1)_X$ model.

{\bf Conclusion.--} We proposed three models to explain the XENON1T excess.
In our models, the $g_{ae}$ couplings are generated by integrating out the vector-like leptons,
and the correct PNGB mass arises from high-dimensional operators.
In the KSVZ axion model, the coupling $g_{a\gamma}$ can be suppressed by choosing
proper sets of vector-like fermions, but we need some fine-tuning to obtain the ALP mass.
In the  $Z_8$  model, the coupling  $g_{a\gamma}$  is suppressed 
due to $SU(3)_C \times U(1)_{\rm EM}$ anomaly free, and the PNGB mass can 
be generated naturally at the keV scale via the dimension-8 operator. 
To solve the PNGB quality problem in the $ Z_8$ model,
we embedded it into the model with $ U(1)_X$ gauge symmetry.

{\bf Acknowledgments.--} This research is supported in part by 
the Projects 11875062 and 11947302 supported by the
National Natural Science Foundation of China, and by
the Key Research Program of Frontier Science, CAS.

\bibliography{refs}

\end{document}